\documentclass[11pt]{article}
\usepackage[dvips]{graphicx}
\usepackage{mathrsfs}
\usepackage{amssymb}
\usepackage{latexsym}
\title{Stackelberg Contention Games in Multiuser Networks}
\author{Jaeok Park\footnote{Department of Economics, University of
California, Los Angeles (UCLA), Los Angeles, CA 90095-1477, USA
(e-mail: jpark31@ucla.edu)} and Mihaela van der
Schaar\footnote{Department of Electrical Engineering, University of
California, Los Angeles (UCLA), Los Angeles, CA 90095-1594, USA
(e-mail: mihaela@ee.ucla.edu)}}
\date{}

\begin{document}

\maketitle

\vspace{-10mm}

\begin{abstract}
Interactions among selfish users sharing a common transmission
channel can be modeled as a non-cooperative game using the game
theory framework. When selfish users choose their transmission
probabilities independently without any coordination mechanism, Nash
equilibria usually result in a network collapse. We propose a
methodology that transforms the non-cooperative game into a
Stackelberg game. Stackelberg equilibria of the Stackelberg game can
overcome the deficiency of the Nash equilibria of the original game.
A particular type of Stackelberg intervention is constructed to show
that any positive payoff profile feasible with independent
transmission probabilities can be achieved as a Stackelberg
equilibrium payoff profile. We discuss criteria to select an
operating point of the network and informational requirements for
the Stackelberg game. We relax the requirements and examine the
effects of relaxation on performance.
\end{abstract}

\section{Introduction}
In wireless communication networks, multiple users often share a
common channel and contend for access. To resolve the contention
problem, many different medium access control (MAC) protocols have
been devised and used. Recently, the selfish behavior of users in
MAC protocols has been studied using game theory. There have been
attempts to understand the existing MAC protocols as the local
utility maximizing behavior of selfish users by reverse-engineering
the current protocols (e.g., \cite{lee}). It has also been
investigated whether existing protocols are vulnerable to the
existence of selfish users who pursue their self-interest in a
non-cooperative manner. Non-cooperative behavior often leads to
inefficient outcomes. For example, in the 802.11 distributed MAC
protocol, DCF, and its enhanced version, EDCF, competition among
selfish users can lead to an inefficient use of the shared channel
in Nash equilibria \cite{tan}. Similarly, a prisoner's dilemma
phenomenon arises in a non-cooperative game for a generalized
version of slotted-Aloha protocols \cite{ma}.

In general, if a game has Nash equilibria yielding low payoffs for
the players, it will be desirable for them to transform the game to
extend the set of equilibria to include better outcomes \cite{myer}.
The same idea can be applied to the game played by selfish users who
compete for access to a common medium. If competition among selfish
users brings about a network collapse, then it is beneficial for
them to design a device which provides incentives to behave
cooperatively. Game theory \cite{myer} discusses three types of
transformation: 1) games with contracts, 2) games with
communication, and 3) repeated games.

A game is said to be with contracts if the players of the game can
communicate and bargain with each other, and enforce the agreement
with a binding contract. The main obstacle to apply this approach to
wireless networking is the distributed nature of wireless networks.
To reach an agreement, users should know the network system and be
able to communicate with each other. They should also be able to
enforce the agreed plan.

A game with communication is the one in which players can
communicate with each other through a mediator but they cannot write
a binding contract. In this case, a correlated equilibrium is
predicted to be played. \cite{altm} studies correlated equilibria
using a coordination mechanism in a slotted Aloha-type scenario.
Unlike the first approach, this does not require that the actions of
players be enforceable. However, to apply this approach to the
medium access problem, signals need to be conveyed from a mediator
to all users, and users need to know the correct meanings of the
signals.

A repeated game is a dynamic game in which the same game is played
repeatedly by the same players over finite or infinite periods.
Repeated interactions among the same players enable them to sustain
cooperation by punishing deviations in subsequent periods. A main
challenge of applying the idea of repeated games to wireless
networks is that the users should keep track of their past
observations and be able to detect deviations and to coordinate
their actions in order to punish deviating users.

Besides the three approaches above, another approach widely applied
to communication networks is pricing \cite{varian}. A central entity
charges prices to users in order to control their utilization of the
network. Nash equilibria with pricing schemes in an Aloha network
are analyzed in \cite{jin,wang}. Implementing a pricing scheme
requires the central entity to have relevant system information as
well as users' benefits and costs, which are often their private
information. Eliciting private information often results in an
efficiency loss in the presence of the strategic behavior of users
as shown in \cite{johari}. Even in the case where the entity has all
the relevant information, prices need to be computed and
communicated to the users.

In this paper, we propose yet another approach using a Stackelberg
game. We introduce a network manager as an additional user and make
him access the medium according to a certain rule. Unlike the
Stackelberg game of \cite{kor} in which the manager (the leader)
chooses a certain strategy before users (followers) make their
decisions, in the proposed Stackelberg game he sets an intervention
rule first and then implements his intervention after users choose
their strategies. Alternatively, the proposed Stackelberg game can
be considered as a generalized Stackelberg game in which there are
multiple leaders (users) and a single follower (the manager) and the
leaders know the response of the follower to their decisions
correctly. With appropriate choices of intervention rules, the
manager can shape the incentives of users in such a way that their
selfish behavior results in cooperative outcomes.

In the context of cognitive radio networks, \cite{bloem} proposes a
related Stackelberg game in which the owner of a licensed frequency
band (the leader) can charge a virtual price for using the frequency
band to cognitive radios (followers). The virtual price signals the
extent to which cognitive radios can exploit the licensed frequency
band. However, since prices are virtual, selfish users may ignore
prices when they make decisions if they can gain by doing so. On the
contrary, in the Stackelberg game of this paper, the intervention of
the manager is not virtual but it results in the reduction of
throughput, which selfish users care about for sure. Hence, the
intervention method provides better grounds for the network manager
to deal with the selfish behavior of users.

\cite{chen1} and \cite{chen2} use game theoretic models to study
random access. Their approach is to capture the information and
implementation constraints using the game theoretic framework and to
specify utility functions so that a desired operating point is
achieved at a Nash equilibrium. If conditions under which a certain
type of dynamic adjustment play converges to the Nash equilibrium
are met, such a strategy update mechanism can be used to derive a
distributed algorithm that converges to the desired operating point.
However, this control-theoretic approach to game theory assumes that
users are obedient. In this paper, our main concern is about the
selfish behavior of users who have innate objectives. Because we
start from natural utility functions and affect them by devising an
intervention scheme, we are in a better position to deal with
selfish users. Furthermore, the idea of intervention can potentially
lead to a distributed algorithm to achieve a desired operating
point.

By formulating the medium access problem as a non-cooperative game,
we show the following main results:
\begin{enumerate}
\item Because the Nash equilibria of the non-cooperative game are inefficient
and/or unfair, we transform the original game into a Stackelberg
game, in which any feasible outcome with independent transmission
probabilities can be achieved as a Stackelberg equilibrium.
\item A particular form of a Stackelberg intervention strategy,
called total relative deviation (TRD)-based intervention, is
constructed and used to achieve any feasible outcome with
independent transmission probabilities.
\item The additional amount of information flows required for the transformation
is relatively moderate, and it can be further reduced without large
efficiency losses.
\end{enumerate}

The rest of this paper is organized as follows. Section 2 introduces
the model and formulates it as a non-cooperative game called the
contention game. Nash equilibria of the contention game are
characterized, and it is shown that they typically yield suboptimal
performance. In Section 3, we transform the contention game into
another related game called the Stackelberg contention game by
introducing an intervening manager. We show that the manager can
implement any transmission probability profile as a Stackelberg
equilibrium using a class of intervention functions. Section 4
discusses natural candidates for the target transmission probability
profile selected by the manager. In Section 5, we discuss the flows
of information required for our results and examine the implications
of some relaxations of the requirements on performance. Section 6
provides numerical results, and Section 7 concludes the paper.

\section{Contention Game Model}
We consider a simple contention model in which multiple users share
a communication channel as in \cite{hamed}. A user represents a
transmitter-receiver pair. Time is divided into slots of the same
duration. Every user has a packet to transmit and can send the
packet or wait. If there is only one transmission, the packet is
successfully transmitted within the time slot. If more than one user
transmits a packet simultaneously in a slot, a collision occurs and
no packet is transmitted.

We summarize the assumptions of our contention model.
\begin{enumerate}
\item A fixed set of users interacts over a given period of time (or
a session).
\item Time is divided into multiple slots, and slots are
synchronized.
\item A user always has a packet to transmit in
every slot.
\item The transmission of a
packet is completed within a slot.
\item A user transmits its packet with the same probability in
every slot. There is no adjustment in the transmission probabilities
during the session. This excludes coordination among users, for
example, using time division multiplexing.
\item There is no cost of transmitting a packet.
\end{enumerate}

We formulate the medium access problem as a non-cooperative game to
analyze the behavior of selfish users. We denote the set of users by
$N = \{1, \ldots, n \}$. Because we assume that a user uses the same
transmission probability over the entire session, the strategy of a
user is its transmission probability, and we denote the strategy of
user $i$ by $p_i$ and the strategy space of user $i$ by $P_i =
[0,1]$ for all $i \in N$.

Once the users decide their transmission probabilities, a strategy
profile can be constructed. The users transmit their packets
independently according to their transmission probabilities, and
thus the strategy profile determines the probability of a successful
transmission by user $i$ in a slot. A strategy profile can be
written as a vector $\mathbf{p} = (p_1, \ldots, p_n)$ in $P = P_1
\times \cdots \times P_n$, the set of strategy profiles. The payoff
function of user $i$, $u_i : P \rightarrow \mathbb{R}$, is defined
as
\begin{eqnarray}
u_i(\mathbf{p}) = k_i p_i \prod_{j \neq i} (1-p_j),
\end{eqnarray}
where $k_i > 0$ measures the value of transmission of user $i$ and
$p_i \prod_{j \neq i} (1-p_j)$ is the probability of successful
transmission by user $i$.

We define the \emph{contention game} by the tuple $\Gamma = \langle
N, (P_i), (u_i) \rangle$. If the users choose their transmission
probabilities taking others' transmission probabilities as given,
then the resulting outcome can be described by the solution concept
of Nash equilibrium \cite{myer}. We first characterize the Nash
equilibria of the contention game.
\newtheorem{prop1}{Proposition}
\begin{prop1}
A strategy profile $\mathbf{p} \in P$ is a Nash equilibrium of the
contention game $\Gamma$ if and only if $p_i = 1$ for at least one
$i$.
\end{prop1}

\noindent\emph{\textbf{Proof}}: In the contention game, the best
response correspondence of user $i$ assumes two sets:
$b_i(\mathbf{p}_{-i}) = \{ 1 \}$ if $\prod_{j \neq i} (1-p_j)>0$ and
$b_i(\mathbf{p}_{-i}) = [0,1]$ if $\prod_{j \neq i} (1-p_j)=0$.
Suppose that user $i$ chooses $p_i = 1$. Then it is playing its best
response while other users are also playing their best responses,
which establishes the sufficiency part. To prove the necessity part,
suppose that $\mathbf{p}$ is a Nash equilibrium and $p_i < 1$ for
all $i \in N$. Since $\prod_{j \neq i} (1-p_j)>0$, $p_i$ is not a
best response to $\mathbf{p}_{-i}$, which is a
contradiction.{\hspace{\stretch{1}} \rule{1ex}{1ex}}

\medskip
If a Nash equilibrium $\mathbf{p}$ has only one user $i$ such that
$p_i = 1$, then $u_i(\mathbf{p}) > 0$ and $u_j(\mathbf{p}) = 0$ for
all $j \neq i$ where $u_i(\mathbf{p})$ can be as large as $k_i$. If
there are at least two users with the transmission probability equal
to 1, then we have $u_i(\mathbf{p}) = 0$ for all $i \in N$. Let
$\mathcal{U}_i = \{ \mathbf{u} \in \mathbb{R}^n : u_i \in [0,k_i],
u_j = 0 \; \forall j \neq i \}$. Then, the set of Nash equilibrium
payoffs is given by
\begin{eqnarray}
\mathcal{U}(NE) = \bigcup_{i=1}^n \mathcal{U}_i.
\end{eqnarray}

Given the game $\Gamma$, we can define the \emph{set of feasible
payoffs} by
\begin{eqnarray} \label{eq:pay}
\mathcal{U} = \{ (u_1(\mathbf{p}), \ldots, u_n(\mathbf{p})) :
\mathbf{p} \in P \}.
\end{eqnarray}
A payoff profile $\mathbf{u}$ in $\mathcal{U}$ is \emph{Pareto
efficient} if there is no other element $\mathbf{v}$ in
$\mathcal{U}$ such that $\mathbf{v} \geq \mathbf{u}$ and $v_i > u_i$
for at least one user $i$. We also call a strategy profile
$\mathbf{p}$ Pareto efficient if $\mathbf{u}(\mathbf{p}) =
(u_1(\mathbf{p}), \ldots, u_n(\mathbf{p}))$ is a Pareto efficient
payoff profile. Let $\mathcal{U}(PE)$ be the set of Pareto efficient
payoffs.

There are $n$ points in $\mathcal{U}(NE) \cap \mathcal{U}(PE)$,
namely, $\mathbf{u}$ such that $u_i = k_i$ and $u_j = 0$ for all $j
\neq i$, for $i = 1, \ldots, n$. These are the corner points of
$\mathcal{U}(PE)$ in which only one user receives a positive payoff.
Therefore, Nash equilibrium payoff profiles are either inefficient
or unfair. Moreover, since $p_i = 1$ is a \emph{weakly dominant
strategy} for every user $i$, in a sense that
$u_i(1,\mathbf{p}_{-i}) \geq u_i(\mathbf{p})$ for all $\mathbf{p}
\in P$, the most likely Nash equilibrium is the one in which $p_i =
1$ for all $i \in N$. At the most likely Nash equilibrium, every
user always transmits its packet, and as a result no packet is
successfully transmitted. Hence, the selfish behavior of the users
is likely to lead to a network collapse, which gives zero payoff to
every user, as argued also in \cite{caga}.

\begin{figure}
\begin{center}
\includegraphics[width=1\textwidth]{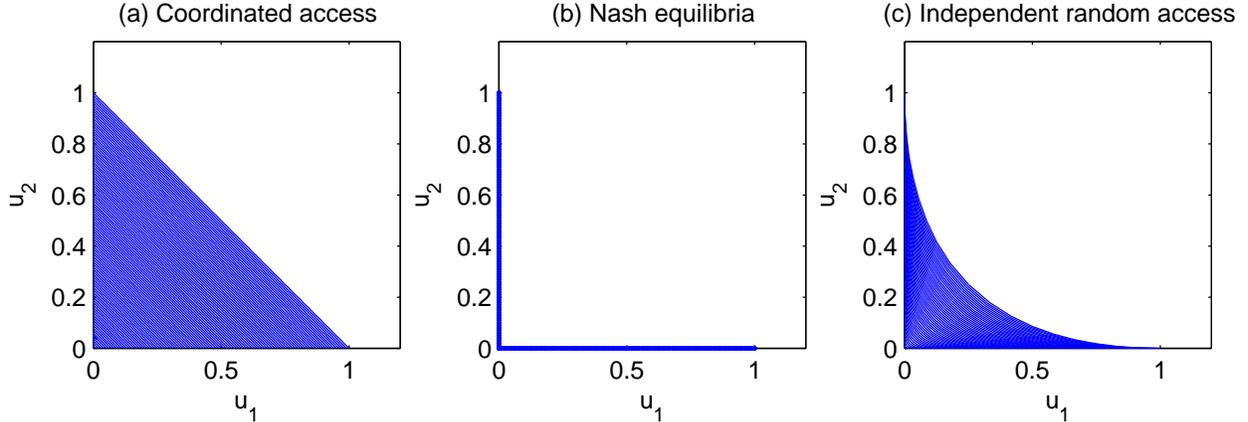}
\end{center}
\caption{Payoff profiles with two homogeneous users with $k_1 = k_2
= 1$. (a) The set of feasible payoffs when coordination between two
users is possible. (b) The set of Nash equilibrium payoffs. (c) The
set of feasible payoffs with independent transmission
probabilities.}
\end{figure}

Figure 1 presents the payoff spaces of two homogeneous users with
$k_1 = k_2 = 1$. If coordination between the two users is possible,
they can achieve any payoff profile in the dark area of Figure 1(a).
For example, $(1/2,1/2)$ can be achieved by arranging user 1 to
transmit only in odd-numbered slots and user 2 only in even-numbered
slots. This kind of coordination can be supported through direct
communications among the users or mediated communications. However,
if such coordination is not possible and each user has to choose one
transmission probability, Nash equilibria yield the payoff profiles
in Figure 1(b). The set of feasible payoffs of the contention game
is shown as the dark area of Figure 1(c). The set of
Pareto-efficient payoff profiles is the frontier of that area. The
lack of coordination makes the set of feasible payoffs smaller
reducing the area of Figure 1(a) to that of Figure 1(c). Because the
typical Nash equilibrium payoff is $(0,0)$, the next section
develops a transformation of the contention game, and the set of
equilibria of the resulting Stackelberg game is shown to expand to
the entire area of Figure 1(c).

\section{Stackelberg Contention Game}
We introduce a network manager as a special kind of user in the
contention game and call him user 0. As a user, the manager can
access the channel with a certain transmission probability. However,
the manager is different from the users in that he can choose his
transmission probability depending on the transmission probabilities
of the users. This ability of the manager enables him to act as the
police. If the users access the channel excessively, the manager can
intervene and punish them by choosing a high transmission
probability, thus reducing the success rates of the users.

Formally, the strategy of the manager is an \emph{intervention
function} $g:P \rightarrow [0,1]$, which gives his transmission
probability $p_0 = g(\mathbf{p})$ when the strategy profile of the
users is $\mathbf{p}$. $g(\mathbf{p})$ can be interpreted as the
level of intervention or punishment by the manager when the users
choose $\mathbf{p}$. Note that the level of intervention by the
manager is the same for every user. We assume that the manager has a
specific ``target'' strategy profile $\tilde{\mathbf{p}}$, that his
transmission has no value to him (as well as to others), and that he
is benevolent. One representation of his objective is the payoff
function of the following form:
\begin{eqnarray}
u_0(g, \mathbf{p}) = \left\{ \begin{array}{ll}
1 - g(\mathbf{p}) & \textrm{if $\mathbf{p} = \tilde{\mathbf{p}}$},\\
0 & \textrm{otherwise}.
\end{array} \right.
\end{eqnarray}
This payoff function means that the manager wants the users to
operate at the target strategy profile $\tilde{\mathbf{p}}$ with the
minimum level of intervention.

We call the transformed game the \emph{Stackelberg contention game}
because the manager chooses his strategy $g$ before the users make
their decisions on the transmission probabilities. In this sense,
the manager can be thought of as a Stackelberg leader and the users
as followers. The specific timing of the Stackelberg contention game
can be outlined as follows:
\begin{enumerate}
\item The network manager determines his intervention function.

\item Knowing the intervention function of the manager, the users
choose their transmission probabilities simultaneously.

\item Observing the strategy profile of the users, the manager
determines the level of intervention using his intervention
function.

\item The transmission probabilities of the manager and the users
determine their payoffs.
\end{enumerate}

Timing 1 happens before the session starts. Timing 2 occurs at the
beginning of the session whereas timing 3 occurs when the manager
knows the transmission probabilities of all the users. Therefore,
there is a time lag between the time when the session begins and
when the manager starts to intervene. Payoffs can be calculated as
the probability of successful transmission averaged over the entire
session, multiplied by valuation. If the interval between timing 2
and timing 3 is short relative to the duration of the session, the
payoff of user $i$ can be approximated as the payoff during the
intervention using the following payoff function:
\begin{eqnarray}
u_i(g, \mathbf{p}) = k_i p_i (1 - g(\mathbf{p})) \prod_{j \neq i}
(1-p_j).
\end{eqnarray}

\begin{figure}
\begin{center}
\includegraphics[width=0.7\textwidth]{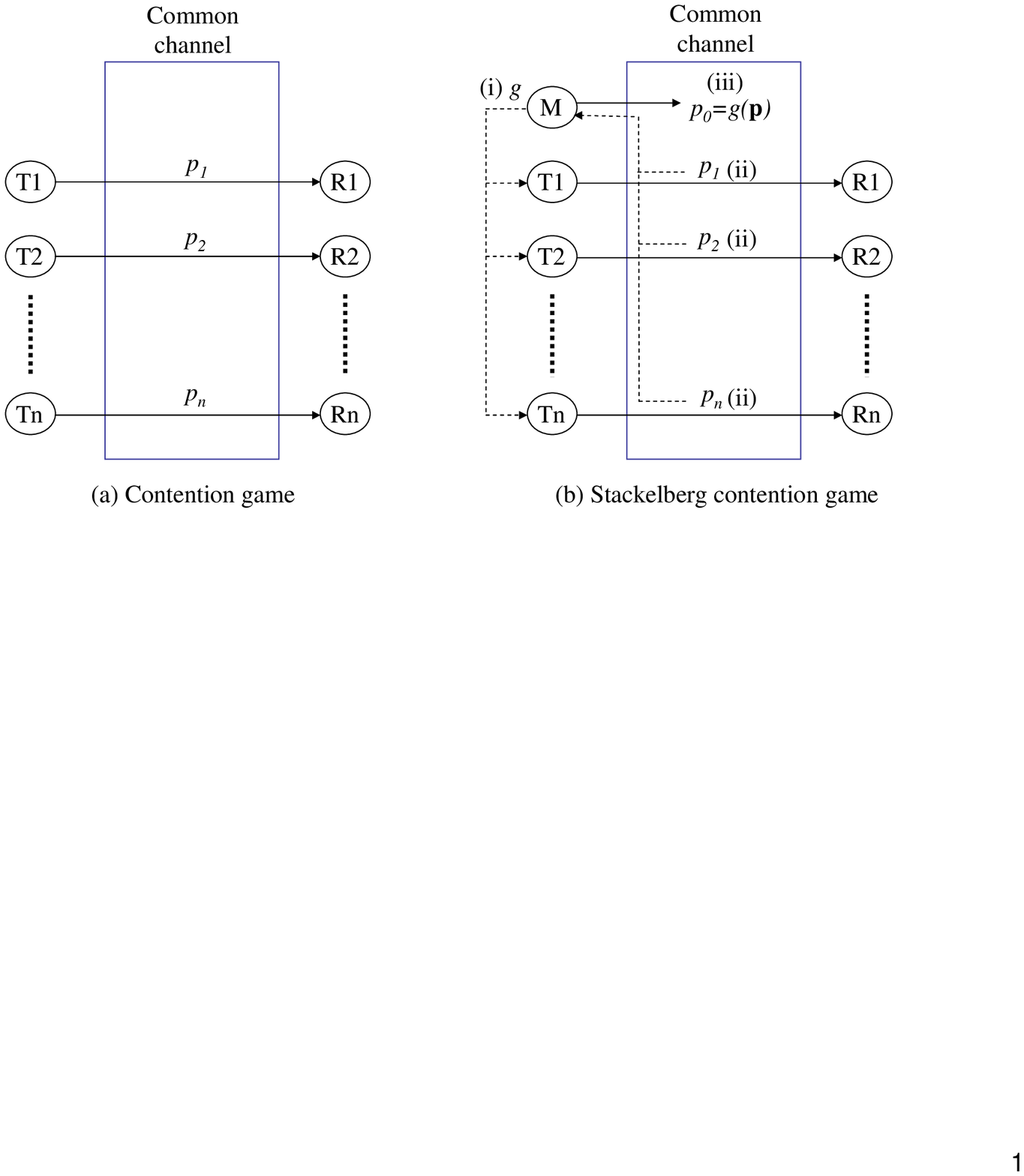}
\end{center}
\caption{Schematic illustration of (a) the contention game and (b)
the Stackelberg contention game. (i),(ii), and (iii) represent the
order of moves in the Stackelberg contention game, and the dotted
arrows represent the flows of information required for the
Stackelberg contention game.}
\end{figure}

The transformation of the contention game into the Stackelberg
contention game is schematically shown in Figure 2. The figure shows
that the main role of the manager is to set the intervention rule
and to implement it. The users still behave non-cooperatively
maximizing their payoffs, and the intervention of the manager
affects their selfish behavior even though the manager does neither
directly control their behavior nor continuously communicate with
the users to convey coordination or price signals.

In the Stackelberg routing game of \cite{kor}, the strategy spaces
of the manager and a user coincide. If that is the case in the
Stackelberg contention game, i.e., if the manager chooses a single
transmission probability before the users choose theirs, then this
intervention only makes the channel lossy but it does not provide
incentives for users not to choose the maximum possible transmission
probability. Hence, in order to provide an incentive to choose a
smaller transmission probability, the manager needs to vary his
transmission probability depending on the transmission probabilities
of the users.

A Stackelberg game is analyzed using a backward induction argument.
The leader predicts the Nash equilibrium behavior of the followers
given his strategy and chooses the best strategy for him. The same
argument can be applied to the Stackelberg contention game. Once the
manager decides his strategy $g$ and commits to implement his
transmission probability according to $g$, the rest of the
Stackelberg contention game (timing 2--4) can be viewed as a
non-cooperative game played by the users. Given the intervention
function $g$, the payoff function of user $i$ can be written as
\begin{eqnarray}
\tilde{u}_i(\mathbf{p};g) = k_i p_i (1 - g(\mathbf{p})) \prod_{j
\neq i} (1-p_j).
\end{eqnarray}
In essence, the role of the manager is to change the non-cooperative
game that the users play from the contention game $\Gamma$ to a new
game $\Gamma_g = \langle N, (P_i), (\tilde{u}_i(\cdot;g)) \rangle$,
which we call the \emph{contention game with intervention $g$}.
Understanding the non-cooperative behavior of the users given the
intervention function $g$, the manager will choose $g$ that
maximizes his payoff.

We now define an equilibrium concept for the Stackelberg contention
game.
\newtheorem{def1}{Definition}
\begin{def1}
An intervention function of the manager $g$ and a profile of the
transmission probabilities of the users $\hat{\mathbf{p}} =
(\hat{p}_1, \ldots, \hat{p}_n)$ constitutes a \emph{Stackelberg
equilibrium} if (i) $\hat{\mathbf{p}}$ is a Nash equilibrium of the
contention game with intervention $g$ and (ii) $\hat{\textbf{p}} =
\tilde{\textbf{p}}$ and $g(\hat{\mathbf{p}}) = 0$.
\end{def1}
Combining (i) and (ii), an equivalent definition is that ($g,
\tilde{\mathbf{p}})$ is a Stackelberg equilibrium if
$\tilde{\mathbf{p}}$ is a Nash equilibrium of $\Gamma_g$ and
$g(\tilde{\mathbf{p}}) = 0$. Condition (i) says that once the
manager chooses his strategy, the users will play a Nash equilibrium
strategy profile in the resulting game, and condition (ii) says that
expecting the Nash equilibrium strategy profile of the users, the
manager chooses his strategy that achieves his objective.

\subsection{Stackelberg Equilibrium with TRD-based Intervention}
As we have mentioned earlier, the manager can choose only one level
of intervention that affects the users equally. A question that
arises is which strategy profile the manager can implement as a
Stackelberg equilibrium with one level of intervention for every
user. We answer this question constructively. We propose a specific
form of an intervention function with which the manager can attain
any strategy profile $\tilde{\mathbf{p}}$ with $0 < \tilde{p}_i < 1$
for all $i$. The basic idea of this result is that because the
strategy of the manager is not a single intervention level but a
function whose value depends on the strategies of the users, he can
discriminate the users by reacting differently to their transmission
probabilities in choosing the level of intervention. Therefore, even
though the realized level of intervention is the same for every
user, the manager can induce the users to choose different
transmission probabilities.

To construct such an intervention function, we first define the
\emph{total relative deviation (TRD) of $\mathbf{p}$ from
$\tilde{\mathbf{p}}$} by
\begin{eqnarray}
h(\mathbf{p}) = \sum_{i = 1}^n \frac{p_i - \tilde{p}_i}{\tilde{p}_i}
= \frac{p_1}{\tilde{p}_1} + \cdots + \frac{p_n}{\tilde{p}_n} - n.
\end{eqnarray}
Since $g$ determines the transmission probability of the manager,
its range should lie in $[0,1]$. To satisfy this constraint, we
define the \emph{TRD-based intervention function} by
\begin{eqnarray}
g^*(\mathbf{p}) = \left[ h(\mathbf{p})\right]_0^1
\end{eqnarray}
where the operator $[x]_a^b = \min \{ \max\{x,a\}, b\}$ is used to
obtain the ``trimmed'' value of TRD between 0 and 1.

The TRD-based intervention can be interpreted in the following way.
The manager sets the target at $\tilde{\mathbf{p}}$. As long as the
users choose small transmission probabilities so that the TRD of
$\mathbf{p}$ from $\tilde{\mathbf{p}}$ does not exceed zero, the
manager does not intervene. If it is larger than zero, the manager
will respond to a one-unit increase in $p_i$ by increasing $p_0$ by
$\frac{1}{\tilde{p}_i}$ units until the TRD reaches 1. The manager
determines the degree of punishment based on the target transmission
probability profile. If he wants a user to transmit with a low
probability, then his punishment against its deviation is strong.

\newtheorem{prop11}[prop1]{Proposition}
\begin{prop11}
$(g^*, \tilde{\mathbf{p}})$ constitutes a Stackelberg equilibrium.
\end{prop11}

\noindent\emph{\textbf{Proof}}: We need to check two things. First,
$\tilde{\mathbf{p}}$ is a Nash Equilibrium of $\Gamma_{g^*}$.
Second, $g^*(\tilde{\mathbf{p}}) = 0$. It is straightforward to
confirm the second. To show the first, the payoff function of user
$i$ given others' strategies $\tilde{\mathbf{p}}_{-i}$ is
\begin{eqnarray}
\tilde{u}_i(p_i, \tilde{\mathbf{p}}_{-i};g^*) &=& k_i p_i \left(1 -
g^*(p_i,\tilde{\mathbf{p}}_{-i}) \right) \prod_{j \neq i} (1-\tilde{p}_j)\\
&=& \left\{ \begin{array}{ll}
0 & \textrm{if $p_i > 2 \tilde{p}_i$},\\
k_i p_i \left(2 - \frac{p_i}{\tilde{p}_i} \right) \prod_{j \neq i}
(1-\tilde{p}_j) & \textrm{if $\tilde{p}_i \leq p_i \leq 2 \tilde{p}_i$},\\
k_i p_i \prod_{j \neq i} (1-\tilde{p}_j) & \textrm{if $p_i <
\tilde{p}_i$}.
\end{array} \right.
\end{eqnarray}
It can be seen from the above expression that $\tilde{u}_i(p_i,
\tilde{\mathbf{p}}_{-i};g^*)$ is increasing on $p_i < \tilde{p}_i$,
reaches a peak at $p_i = \tilde{p}_i$, is decreasing on $\tilde{p}_i
< p_i < 2 \tilde{p}_i$, and then stays at 0 on $p_i \geq 2
\tilde{p}_i$. Therefore, user $i$'s best response to
$\tilde{\mathbf{p}}_{-i}$ is $\tilde{p}_i$ for all $i$, and thus
$\tilde{\mathbf{p}}$ constitutes a Nash Equilibrium of the
contention game with TRD-based intervention,
$\Gamma_{g^*}$.{\hspace{\stretch{1}} \rule{1ex}{1ex}}

\newtheorem{corr1}{Corollary}
\begin{corr1}
Any feasible payoff profile $\mathbf{u} \in \mathcal{U}$ of the
contention game with $u_i > 0$ for all $i \in N$ can be achieved by
a Stackelberg equilibrium.
\end{corr1}

Corollary 1 resembles the Folk theorem of repeated games \cite{myer}
in that it claims that any feasible outcome can be attained as an
equilibrium. Incentives not to deviate from a certain operating
point are provided by the manager's intervention in the Stackelberg
contention game, while in a repeated game players do not deviate
since a deviation is followed by punishment from other players.

\subsection{Nash Equilibria of the Contention Game with TRD-based Intervention}
In Proposition 2, we have seen that $\tilde{\mathbf{p}}$ is a Nash
equilibrium of the contention game with TRD-based intervention.
However, if other Nash equilibria exist, the outcome may be
different from the one that the manager intends. In fact, any
strategy profile $\mathbf{p}$ with $p_i = 1$ for at least one $i$ is
still a Nash equilibrium of $\Gamma_{g^*}$. The following
proposition characterizes the set of Nash equilibria of
$\Gamma_{g^*}$ that are different from those of $\Gamma$.

\newtheorem{prop2}[prop1]{Proposition}
\begin{prop2}
Consider a strategy profile $\hat{\mathbf{p}}$ with $\hat{p}_i < 1$
for all $i \in N$. $\hat{\mathbf{p}}$ is a Nash equilibrium of the
contention game with TRD-based intervention if and only if either
\begin{eqnarray}
(i)\quad \hat{\mathbf{p}} = \tilde{\mathbf{p}}
\end{eqnarray}
or
\begin{eqnarray}
(ii)\quad \sum_{j \neq i} \frac{\hat{p}_j -
\tilde{p}_j}{\tilde{p}_j} \geq 2 \textrm{ for all $i = 1, \ldots,
n$}.
\end{eqnarray}
\end{prop2}

\noindent\emph{\textbf{Proof}}: See Appendix A.
{\hspace{\stretch{1}} \rule{1ex}{1ex}}

\medskip
Transforming $\Gamma$ to $\Gamma_{g^*}$ does not eliminate the Nash
equilibria of the contention game. Rather, the set of Nash
equilibria expands to include two classes of new equilibria. The
first Nash equilibrium of Proposition 3 is the one that the manager
intends the users to play. The second class of Nash equilibria are
those in which the sum of relative deviations of other users is
already too large that no matter how small transmission probability
user $i$ chooses, the level of intervention stays the same at 1.

Since $\tilde{\mathbf{p}}$ is chosen to satisfy $0 < \tilde{p}_i <
1$ for all $i$ and $g^*$ satisfies $g^*(\tilde{\mathbf{p}}) = 0$, it
follows that $\tilde{u}_i(\tilde{\mathbf{p}}) > 0$ for all
$i$.\footnote{Since we mostly consider the TRD-based intervention
function $g^*$, we will use $\tilde{u}_i(\tilde{\mathbf{p}})$
instead of $\tilde{u}_i(\tilde{\mathbf{p}};g^*)$ when there is no
confusion.} For the second class of Nash equilibria in Proposition
3, $\tilde{u}_i(\hat{\mathbf{p}}) = 0$ for all $i$ because
$g^*(\hat{\mathbf{p}}) = 1$. Therefore, the payoff profile of the
second class of Nash equilibria is \emph{Pareto dominated} by that
of the intended Nash equilibrium in that the intended Nash
equilibrium yields a higher payoff for every user compared to the
second class of Nash equilibria.

The same conclusion holds for Nash equilibria with more than one
user with transmission probability 1 because every user gets zero
payoff. Finally, the remaining Nash equilibria are those with
exactly one user with transmission probability 1. Suppose that $p_i
= 1$. Then the highest payoff for user $i$ is achieved when $p_j =
0$ for all $j \neq i$. Denoting this strategy profile by
$\mathbf{e}_i$, the payoff profile of $\mathbf{e}_i$ is Pareto
dominated by that of $\tilde{\mathbf{p}}$ if $1 - g^*(\mathbf{e}_i)
= 1 + n - \frac{1}{\tilde{p}_i} < \tilde{p}_i \prod_{j \neq i}
(1-\tilde{p}_j)$.

\subsection{Reaching the Stackelberg Equilibrium}
We have seen that there are multiple Nash equilibria of the
contention game with TRD-based intervention and that the Nash
equilibrium $\tilde{\mathbf{p}}$ in general yields higher payoffs to
the users than other Nash equilibria. If the users are aware of the
welfare properties of different Nash equilibria, they will tend to
select $\tilde{\mathbf{p}}$.

Suppose that the users play the second class of Nash equilibria in
Proposition 3 for some reason. If the Stackelberg contention game is
played repeatedly and the users anticipate that the strategy profile
of the other users will be the same as that of the last period, then
it can be shown that under certain conditions there is a sequence of
intervention functions convergent to $g^*$ that the manager can
employ to have the users reach the intended Nash equilibrium
$\tilde{\mathbf{p}}$, thus approaching the Stackelberg equilibrium.

\newtheorem{prop3}[prop1]{Proposition}
\begin{prop3}
Suppose that at $t = 0$ the manager chooses the intervention
function $g^*$ and that the users play a Nash equilibrium
$\hat{\mathbf{p}}^0$ of the second class.

Without loss of generality, the users are enumerated so that the
following holds:
\begin{eqnarray}
\frac{\hat{p}_1^0}{\tilde{p}_1} \leq \frac{\hat{p}_2^0}{\tilde{p}_2}
\leq \ldots \leq \frac{\hat{p}_{n-1}^0}{\tilde{p}_{n-1}} \leq
\frac{\hat{p}_n^0}{\tilde{p}_n}.
\end{eqnarray}
Suppose further that for each $i$, either
$\frac{\hat{p}_n^0}{\tilde{p}_n} - \frac{\hat{p}_i^0}{\tilde{p}_i} <
2$ or $\frac{\hat{p}_i^0}{\tilde{p}_i} \leq 1$ holds.

At $t \geq 1$; Define
\begin{eqnarray}
c^t = \sum_{j \neq n} \frac{\hat{p}_j^{t-1}}{\tilde{p}_j} + 1.
\end{eqnarray}
Assume that the manager employs the intervention function
$g^t(\mathbf{p}) = [h^t(\mathbf{p})]_0^1$ where
\begin{eqnarray}
h^t(\mathbf{p}) = \frac{p_1}{\tilde{p}_1} + \cdots +
\frac{p_n}{\tilde{p}_n} - c^t
\end{eqnarray}
and that user $i$ chooses $\hat{p}_i^t$ as a best response to
$\hat{\mathbf{p}}_{-i}^{t-1}$ given $g^t$.

Then $\lim_{t \rightarrow \infty} \hat{p}_i^t = \tilde{p}_i$ for all
$i = 1, \ldots, n$ and $\lim_{t \rightarrow \infty} c^t = n$.
\end{prop3}

\noindent\emph{\textbf{Proof}}: See Appendix B.
{\hspace{\stretch{1}} \rule{1ex}{1ex}}

\medskip
The reason that no user has an incentive to deviate from the second
class of Nash equilibria is that since others use high transmission
probabilities, the TRD is over 1 no matter what transmission
probability a user chooses. Since the punishment level is always 1,
a reduction of the transmission probability by a user is not
rewarded by a decreased level of intervention. If the relative
deviations of $p_i$ from $\tilde{p}_i$ are not too disperse, the
manager can successively adjust down the effective range of
punishment so that he can react to the changes in the strategies of
the users. Proposition 4 shows that this procedure succeeds to have
the strategy profile of the users converge to the intended Nash
equilibrium.

\section{Target Selection Criteria of the Manager}
So far we have assumed that the manager has a target strategy
profile $\tilde{\mathbf{p}}$ and examined whether he can find an
intervention function that implements it as a Stackelberg
equilibrium. This section discusses selection criteria that the
manager can use to choose the target strategy profile. To address
this issue, we rely on cooperative game theory because a reasonable
choice of the manager should have a close relationship to the likely
outcome of bargaining among the users if bargaining were possible
for them \cite{myer}. The absence of communication opportunities
among the users prevents them from engaging in bargaining or from
directly coordinating with each other.

\subsection{Nash Bargaining Solution}
The pair $(F, \mathbf{v})$ is an \emph{n-person bargaining problem}
where $F$ is a closed and convex subset of $\mathbb{R}^n$,
representing the set of feasible payoff allocations and $\mathbf{v}
= (v_1, \ldots, v_n)$ is the disagreement payoff allocation. Suppose
that there exists $\mathbf{y} \in F$ such that $y_i > v_i$ for every
$i$.

\newtheorem{def3}[def1]{Definition}
\begin{def3}
$\mathbf{x}$ is the \emph{Nash bargaining solution} for an n-person
bargaining problem $(F,\mathbf{v})$ if it is the unique Pareto
efficient vector that solves
\begin{eqnarray}
\max_{\mathbf{x} \in F, \mathbf{x} \geq \mathbf{v}} \prod_{i=1}^n
(x_i - v_i).
\end{eqnarray}
\end{def3}

Consider the contention game $\Gamma$. $(\mathcal{U},\mathbf{0})$
can be regarded as an $n$-person bargaining problem where
$\mathcal{U}$ is defined in (\ref{eq:pay}) and $\mathbf{0}$ is the
disagreement point. The vector $\mathbf{0}$ is the natural
disagreement point because it is a Nash equilibrium payoff as well
as the minimax value for each user. The only departure from the
standard theory is that the set of feasible payoffs $\mathcal{U}$ is
not convex.\footnote{We do not allow public randomization among
users, which requires coordination among them.} However, we can
carry the definition of the Nash bargaining solution to our setting
as in \cite{caga}.

Since the manager knows the structure of the contention game, he can
calculate the Nash bargaining solution $\mathbf{u}$ for
$(\mathcal{U},\mathbf{0})$ and find the strategy profile
$\tilde{\mathbf{p}}$ that yields $\mathbf{u}$. Then the manager can
implement $\tilde{\mathbf{p}}$ by choosing $g^*$ based on
$\tilde{\mathbf{p}}$. Notice that the presence of the manager does
not decrease the payoffs of the users because
$g^*(\tilde{\mathbf{p}})$ = 0. The Nash bargaining solution for
$(\mathcal{U},\mathbf{0})$ has the following simple form.

\newtheorem{prop4}[prop1]{Proposition}
\begin{prop4}
$\frac{(n-1)^{n-1}}{n^n} (k_1, \ldots, k_n)$ is the Nash bargaining
solution for $(\mathcal{U},\mathbf{0})$, and it is attained by $p_i
= \frac{1}{n}$ for all $i = 1, \ldots, n$.
\end{prop4}

\noindent\emph{\textbf{Proof}}: The maximand in the definition of
the Nash bargaining solution can be written as
\begin{eqnarray}
\max_{\mathbf{u} \in \mathcal{U}, \mathbf{u} \geq \mathbf{0}}
\prod_{i=1}^n u_i.
\end{eqnarray}
Since any $\mathbf{u} \in \mathcal{U}$ satisfies $\mathbf{u} \geq
\mathbf{0}$, the above problem can be expressed in terms of
$\mathbf{p}$:
\begin{eqnarray}
\max_{\mathbf{p} \in P} \left(\prod_{i=1}^n k_i \right)
\prod_{i=1}^n p_i (1-p_i)^{n-1}.
\end{eqnarray}
The logarithm of the objective function is strictly concave in
$\mathbf{p}$, and the first-order optimality condition gives $p_i =
\frac{1}{n}$ for all $i = 1, \ldots, n$. {\hspace{\stretch{1}}
\rule{1ex}{1ex}}

\medskip
The Nash bargaining solution for $(\mathcal{U},\mathbf{0})$ treats
every user equally in that it specifies the same transmission
probability for every user. Therefore, the manager does not need to
know the vector of the values of transmission $\mathbf{k} = (k_1,
\ldots, k_n)$ to implement the Nash bargaining solution. The Nash
bargaining solution coincides with the Kalai-Smorodinsky solution
\cite{kalai} because the maximum payoff for user $i$ is $k_i$ and
the Nash bargaining solution is the unique efficient payoff profile
in which each user receives a payoff proportional to its maximum
feasible payoff.

If the manager wants to treat the users with discrimination, he can
use the \emph{generalized Nash product}
\begin{eqnarray}
\prod_{i=1}^n (x_i - v_i)^{\omega_i}
\end{eqnarray}
as the maximand to find a \emph{nonsymmetric Nash bargaining
solution}, where $\omega_i > 0$ represents the weight for user $i$.
One example of the weights is the valuation of the
users.\footnote{If $k_i$ is private information, it would be
interesting to construct a mechanism that induces users to reveal
their true values $k_i$.} The nonsymmetric Nash bargaining solution
for $(\mathcal{U},\mathbf{0})$ can be shown to be achieved by $p_i =
\frac{\omega_i}{\sum_i \omega_i}$ for all $i$ using the similar
method to the proof of Proposition 5.

\subsection{Coalition-Proof Strategy Profile}
If some of the users can communicate and collude effectively, the
network manager may want to choose a strategy profile which is
self-enforcing even in the existence of coalitions. Since we define
a user as a transmitter-receiver pair, a collusion may occur when a
single transmitter sends packets to several destinations and
controls the transmission probabilities of several users.

Given the set of users $N = \{ 1, \ldots, n \}$, a \emph{coalition}
is any nonempty subset $S$ of $N$. Let $\mathbf{p}_S$ be the
strategy profile of the users in $S$.

\newtheorem{def4}[def1]{Definition}
\begin{def4}
$\tilde{\mathbf{p}}$ is \emph{coalition-proof} with respect to a
coalition $S$ in a non-cooperative game $\langle N,[0,1]^N,(u_i)
\rangle$ if there does not exist $\mathbf{p}_S \in [0,1]^S$ such
that $u_i(\mathbf{p}_S, \tilde{\mathbf{p}}_{-S}) \geq
u_i(\tilde{\mathbf{p}})$ for all $i \in S$ and $u_i(\mathbf{p}_S,
\tilde{\mathbf{p}}_{-S}) > u_i(\tilde{\mathbf{p}})$ for at least one
user $i \in S$.
\end{def4}

By definition, $\tilde{\mathbf{p}}$ is coalition-proof with respect
to the \emph{grand coalition} $S = N$ if and only if
$\mathbf{u}(\tilde{\mathbf{p}}) = ({u}_1(\tilde{\mathbf{p}}),
\ldots, {u}_n(\tilde{\mathbf{p}}))$ is Pareto efficient. If
$\tilde{\mathbf{p}}$ is a Nash equilibrium, then it is
coalition-proof with respect to any one-person ``coalition.'' The
non-cooperative game of our interest is the contention game with
TRD-based intervention $g^*$.

\newtheorem{prop5}[prop1]{Proposition}
\begin{prop5}
$\tilde{\mathbf{p}}$ is coalition-proof with respect to a two-person
coalition $S = \{ i, j \}$ in the contention game with TRD-based
intervention $g^*$ if and only if $\tilde{p}_i + \tilde{p}_j \leq
1$.
\end{prop5}

\noindent\emph{\textbf{Proof}}: See Appendix C.
{\hspace{\stretch{1}} \rule{1ex}{1ex}}

\medskip
The proof of Proposition 6 shows that if $\tilde{p}_i + \tilde{p}_j
> 1$ then users $i$ and $j$ can jointly reduce their transmission
probabilities to increase their payoffs at the same time. For
example, suppose that users 1 and 2 are controlled by the same
transmitter and that the manager selects the target
$\tilde{\mathbf{p}}$ with $\tilde{p}_1 = 0.3$ and $\tilde{p}_2 =
0.8$. Then $\tilde{u}_1(\tilde{\mathbf{p}}) = 0.06 \, k_1 \prod_{j
\neq 1,2} (1 - \tilde{p}_j)$ and $\tilde{u}_2(\tilde{\mathbf{p}}) =
0.56 \, k_2 \prod_{j \neq 1,2} (1 - \tilde{p}_j)$. Suppose that the
two users jointly deviate to $(p_1,p_2) = (0.25, 0.75)$. Then the
new payoffs are $\tilde{u}_1(p_1,p_2,\tilde{\mathbf{p}}_{N \setminus
\{1,2\}}) = 0.0625\, k_1 \prod_{j \neq 1,2} (1 - \tilde{p}_j)$ and
$\tilde{u}_2(p_1,p_2,\tilde{\mathbf{p}}_{N \setminus \{1,2\}}) =
0.5625\, k_2 \prod_{j \neq 1,2} (1 - \tilde{p}_j)$, which is
strictly better for both users. A decrease in $p_i$ and $p_j$ at the
same time also increases the payoffs of all the users not belonging
to the coalition, which implies that a target $\tilde{\mathbf{p}}$
with $\tilde{p}_i + \tilde{p}_j > 1$ is not Pareto efficient. This
observation leads to the following corollary.

\newtheorem{corr2}[corr1]{Corollary}
\begin{corr2}
If $\tilde{\mathbf{p}}$ is Pareto efficient in the contention game
with TRD-based intervention $g^*$, then it is coalition-proof with
respect to any two-person coalition.
\end{corr2}

In fact, we can generalize the above corollary and provide a
stronger statement.
\newtheorem{prop55}[prop1]{Proposition}
\begin{prop55}
$\tilde{\mathbf{p}}$ is Pareto efficient in the contention game with
TRD-based intervention $g^*$ if and only if it is coalition-proof
with respect to any coalition.
\end{prop55}

\noindent\emph{\textbf{Proof}}: See Appendix D.
{\hspace{\stretch{1}} \rule{1ex}{1ex}}

\section{Informational Requirement and Its Relaxation}
We have introduced and analyzed the contention game and the
Stackelberg contention game with TRD-based intervention. In this
section we discuss what the players of each game need to know in
order to play the corresponding equilibrium.

\subsection{Contention Game and Nash Equilibrium}
In a general non-cooperative game, each user needs to know, or
predict correctly, the strategy profile of others in order to find
its best response strategy. In the contention game with the payoff
function $u_i(\mathbf{p}) = k_i p_i \prod_{j \neq i} (1-p_j)$, it
suffices for user $i$ to know the sign of $\prod_{j \neq i}
(1-p_j)$, i.e., whether it is positive or zero, to calculate its
best response. On the other hand, $p_i = 1$ is a \emph{weakly
dominant strategy} for any user $i$, which means setting $p_i = 1$
is weakly better no matter what strategies other users choose.
Hence, the Nash equilibrium $\mathbf{p} = (1,\ldots,1)$ does not
require any knowledge on others' strategies.

\subsection{Stackelberg Contention Game with TRD-based Intervention
and Stackelberg Equilibrium}
Considering the timing of the Stackelberg contention game outlined
in Section 3, we can list the following requirements on the manager
and the users for the Stackelberg equilibrium to be played.

\medskip
\noindent \textbf{Requirement M.} Once the users choose the
transmission probabilities, the manager observes the strategy
profile of the users.
\medskip

The manager needs to decide the level of intervention as a function
of the transmission probabilities of the users. If the manager can
distinguish the access of each user and have sufficiently many
observations to determine the transmission probability of each user,
then this requirement will be satisfied. If the manager can observe
the channel state (idle, success, collision) and identify the users
of successfully transmitted packets, he can estimate the
transmission probability of each user in the following way. First,
he can obtain an estimate of $\prod_{i \in N} (1-p_i)$ by
calculating the frequency of idle slots, called $q_{idle}$. Second,
he can obtain an estimate of $p_i \prod_{j \neq i} (1-p_j)$ by
calculating the frequency of slots in which user $i$ succeeds to
transmit its packet, called $q_{i}$. Finally, an estimate of $p_i$
can be obtained by solving $\frac{p_i}{1-p_i} =
\frac{q_i}{q_{idle}}$ for $p_i$.

\smallskip \smallskip \noindent \textbf{Requirement U.} User $i$ knows
$g^*$ (and thus $\tilde{\mathbf{p}}$) and $\mathbf{p}_{-i}$ when it
chooses its transmission probability.\smallskip \smallskip

Requirement U is sufficient for the Nash equilibrium of the
contention game with TRD-based intervention to be played by the
users. User $i$ can find its best response strategy by maximizing
$\tilde{u}_i$ given $g^*$ and $\mathbf{p}_{-i}$. In fact, a weaker
requirement is compatible with the Nash equilibrium of the
contention game with TRD-based intervention. Suppose that user $i$
knows the \emph{form} of intervention function $g^*$ and the value
of $\tilde{p}_i$, and observes the intervention level $p_0$.
$\tilde{\mathbf{p}}$ embedded in the TRD-based intervention function
$g^*$ can be thought of as a recommended strategy profile by the
manager (thus the communication from the manager to the users occurs
indirectly through the function $g^*$). Even though user $i$ does
not know the recommended strategies to other users, i.e., the values
of $\tilde{p}_j$, $j \neq i$, it knows its recommended transmission
probability. From the form of the intervention function, user $i$
can derive that it is of its best interest to follow the
recommendation as long as all the other users follow their
recommended strategies. Observing $p_0 = 0$ confirms its belief that
other users play the recommended strategies, and it has no reason to
deviate.

The users can acquire knowledge on the intervention function $g^*$
through one of three ways: (i) known protocol, (ii) announcement,
and (iii) learning. The first method is effective in the case where
a certain network manager operates in a certain channel (for
example, a frequency band). The community of users will know the
protocol (or intervention function) used by the manager. This method
does not require any information exchange between the manager and
the users. Neither teaching of the manager nor learning of the users
is necessary. However, there is inflexibility in choosing an
intervention function, and the manager cannot change his target
strategy profile frequently. Nevertheless, this is the method most
often used in current wireless networks, where users appertain to a
predetermined class of known and homogeneous protocols.

The second method allows the manager to make the users know $g^*$
directly, which includes information on the target
$\tilde{\mathbf{p}}$. The manager will execute his intervention
according to the announced intervention function because the
Stackelberg equilibrium $(g^*, \tilde{\mathbf{p}})$ achieves his
objective. However, it requires explicit message delivery from the
manager to the users, which is sometimes costly or may even be
impossible in practice.

Finally, if the Stackelberg contention game is played repeatedly
with the same intervention function, the users may be able to
recover the form of the intervention function chosen by the manager
based on their observations on $(p_0,{\mathbf{p}})$, for example,
using learning techniques developed in \cite{wehu, huwe, voro}.
However, this process may take long and the users may not be able to
collect enough data to find out the true functional form if there is
limited experimentation of the users.

\emph{Remark.} If users are obedient, the manager can use
centralized control by communicating $\tilde{p}_i$ to user $i$.
Additional communication and estimation overhead required for the
Stackelberg equilibrium can be considered as a cost incurred to deal
with the selfish behavior of users, or to provide incentives for
users to follow $\tilde{\mathbf{p}}$.

\subsection{Limited Observability of the Manager}
The construction of the TRD-based intervention function assumes that
the manager can observe or estimate the transmission probabilities
of the users correctly. In real applications, the manager may not be
able to observe the exact choice made by each user. We consider
several scenarios under which the manager has limited observability
and examine how the TRD-based intervention function can be modified
in those scenarios.

\subsubsection{Quantized Observation}
Let $\mathcal{I} = \{I_0, I_1, \ldots, I_m \}$ be a set of intervals
which partition $[0,1]$. We assume that each interval contains its
right end point. For simplicity, we will consider intervals of the
same length. That is, $\mathcal{I} = \left\{ \{0\},
\left(0,\frac{1}{m}\right], \left(\frac{1}{m}, \frac{2}{m}\right],
\ldots, \left(\frac{m-1}{m}, 1 \right] \right\}$, and we call $I_0 =
\{0\}$ and $I_r = \left(\frac{r-1}{m}, \frac{r}{m}\right]$ for all
$r = 1, \ldots, m$.

Suppose that the manager only observes which interval in
$\mathcal{I}$ each $p_i$ belongs to. In other words, the manager
observes $r_i$ instead of $p_i$ such that $p_i \in I_{r_i}$. In this
case, the level of intervention is calculated based on $\mathbf{r} =
(r_1, \ldots, r_n)$ rather than $\mathbf{p}$. It means that given
$\mathbf{p}_{-i}$, $p_0$ would be the same for any $p_i, p'_i$ if
$p_i$ and $p'_i$ belong to the same $I_{r}$. Since any $p_i \in
\left(\frac{r-1}{m}, \frac{r}{m}\right)$ is weakly dominated by $p_i
= \frac{r}{m}$, the users will choose their transmission
probabilities at the right end points of the intervals in
$\mathcal{I}$. This in turn will affect the choice of a target by
the manager. The manager will be restricted to choose
$\tilde{\mathbf{p}}$ such that $\tilde{p}_i \in \left\{ \frac{1}{m},
\ldots, \frac{m-1}{m} \right\}$ for all $i \in N$. Then the manager
can implement $\tilde{\mathbf{p}}$ with the intervention function
$g(\mathbf{r}) = g^*(\mathbf{p})$, where $p_i$ is set equal to
$\frac{r_i}{m}$. In summary, the quantized observation on
$\mathbf{p}$ restricts the choice of $\tilde{\mathbf{p}}$ by the
manager from $(0,1)^N$ to $\left\{ \frac{1}{m}, \ldots,
\frac{m-1}{m} \right\}^N$.

\begin{figure}
\begin{center}
\includegraphics[width=0.7\textwidth]{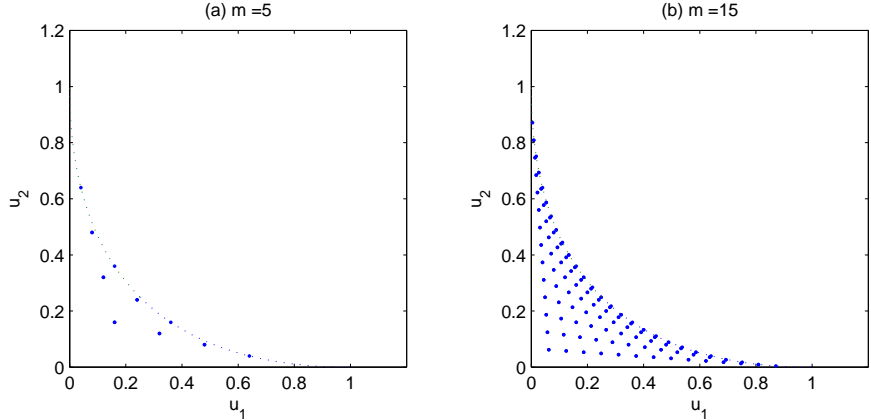}
\end{center}
\caption{Payoffs that can be achieved by the manager with quantized
observation. (a) $m = 5$. (b) $m = 15$. }
\end{figure}

Figure 3 shows the payoff profiles that can be achieved by the
manager with quantized observation. When the number of intervals is
moderately large, the manager has many options near or on the Pareto
efficiency boundary.

\subsubsection{Noisy Observation}
We modify the Stackelberg contention game to analyze the case where
the manager observes noisy signals of the transmission probabilities
of the users. Let $P_i = [\epsilon, 1 - \epsilon]$ be the strategy
space of user $i$, where $\epsilon$ is a small positive number. We
assume that the users can observe the strategy profile $\mathbf{p}$,
but the manager observes a noisy signal of $\mathbf{p}$. The manager
observes $p_i^o$ instead of $p_i$ where $p_i^o$ is uniformly
distributed on $[p_i - \epsilon, p_i + \epsilon]$, independently
over $i \in N$. Suppose that the manager chooses a target
$\tilde{\mathbf{p}}$ such that $\tilde{p}_i \in [2\epsilon, 1 -
2\epsilon]$. The expected payoff of user $i$ when the manager uses
an intervention function $g$ is
\begin{eqnarray}
E[\tilde{u}_i(\mathbf{p};g)|\mathbf{p}] = k_i p_i \prod_{j \neq i}
(1-p_j) \left(1 - E[g(\mathbf{p}^o)|\mathbf{p}] \right).
\end{eqnarray}
Hence, the intervention function is effectively
$E[g(\mathbf{p}^o)|\mathbf{p}]$ instead of $g(\mathbf{p})$ when the
manager observes $\mathbf{p}^o$. If $\hat{\mathbf{p}}$ is a Nash
equilibrium of the contention game with intervention $g$ when
$\mathbf{p}$ is perfectly observable to the manager and
$E[g(\mathbf{p}^o)|\mathbf{p}] = g(\mathbf{p})$ for all $\mathbf{p}$
such that $\max_{i \in N} |p_i - \hat{p}_i | \leq \epsilon$, then
$\hat{\mathbf{p}}$ will still be a Nash equilibrium of the
contention game with intervention $g$ when the manager observes a
noisy signal of the strategy profile of the users.

Consider the TRD-based intervention function $g^*$. Since
$g^*(\mathbf{p}) \geq 0$ for all $\mathbf{p} \in P$ and
$h(\mathbf{p}^o) > 0$ with a positive probability when $\mathbf{p} =
\tilde{\mathbf{p}}$, $E[g^*(\mathbf{p}^o)|\tilde{\mathbf{p}}] > 0$
whereas $g^*(\tilde{\mathbf{p}}) = 0$. Since $g^*$ is kinked at
$\tilde{\mathbf{p}}$, the noise in $\mathbf{p}^o$ will distort the
incentives of the users to choose $\tilde{\mathbf{p}}$.

The manager can implement his target $\tilde{\mathbf{p}}$ at the
expense of intervention with a positive probability. If the manager
adopts the following intervention function
\begin{eqnarray}\label{eq:modtrd}
g(\mathbf{p}) = \sum_{i \in N} \frac{ \frac{1}{1 + \epsilon q } p_i
- \tilde{p}_i}{\tilde{p}_i} + \frac{(n+1) \epsilon q}{1+ \epsilon
q},
\end{eqnarray}
where $q = \sum_{i \in N} \frac{1}{\tilde{p}_i}$, then
$\tilde{\mathbf{p}}$ is a Nash equilibrium of the contention game
with intervention $g$, but the average level of intervention at
$\tilde{\mathbf{p}}$ is
\begin{eqnarray}
E[g(\mathbf{p}^o)|\tilde{\mathbf{p}}] = g(\tilde{\mathbf{p}}) =
\frac{\epsilon q}{1+ \epsilon q} > 0,
\end{eqnarray}
which can be thought of as the efficiency loss due to the noise in
observations.

\begin{figure}
\begin{center}
\includegraphics[width=0.7\textwidth]{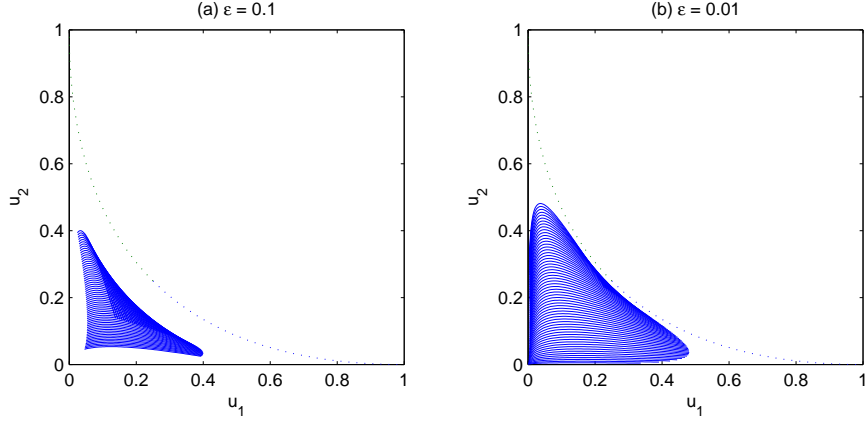}
\end{center}
\caption{Payoffs that can be achieved by the manager with noisy
observation. (a) $\epsilon = 0.1$. (b) $\epsilon = 0.01$. }
\end{figure}

Figure 4 illustrates the set of payoff profiles that can be achieved
with the intervention function given by (\ref{eq:modtrd}). As the
size of the noise gets smaller, the set expands to approach the
Pareto efficiency boundary.

\subsubsection{Observation on the Aggregate Probability}
We consider the case where the manager can observe only the
frequency of the slots that are not accessed by any user. If the
users transmit their packets according to $\mathbf{p}$, then the
manager observes only the aggregate probability $\prod_{i \in N}
(1-p_i)$. In this scenario, the intervention function that the
manager chooses has to be a function of $\prod_{i \in N} (1-p_i)$,
and this implies that the manager cannot discriminate among the
users.

The TRD-based intervention function $g^*$ allows the manager to use
different reactions to each user's deviation. In the effective
region where the TRD is between 0 and 1, one unit increase in $p_i$
results in $\frac{1}{\tilde{p}_i}$ units increase in $p_0$. However,
this kind of discrimination through the structure of the
intervention function is impossible when the manager cannot observe
individual transmission probabilities.

This limitation forces the manager to treat the users equally, and
the target has to be chosen such that $\tilde{p}_i = \tilde{p}$ for
all $i \in N$. If the manager uses the following intervention
function,
\begin{eqnarray}
g(\mathbf{p}) = \left[ \frac{1}{\tilde{p} (1-\tilde{p})^{n-1} }
\left( ( 1 - \tilde{p})^n - \prod_{i \in N} (1-p_i) \right)
\right]_0^1
\end{eqnarray}
then he can implement $\tilde{\mathbf{p}} = (\tilde{p}, \ldots,
\tilde{p})$ with $g(\tilde{\mathbf{p}}) = 0$ as a Stackelberg
equilibrium. Hence, if the manager only observes the aggregate
probability, this prevents him from setting the target transmission
probabilities differently across users.

\begin{figure}
\begin{center}
\includegraphics[width=0.7\textwidth]{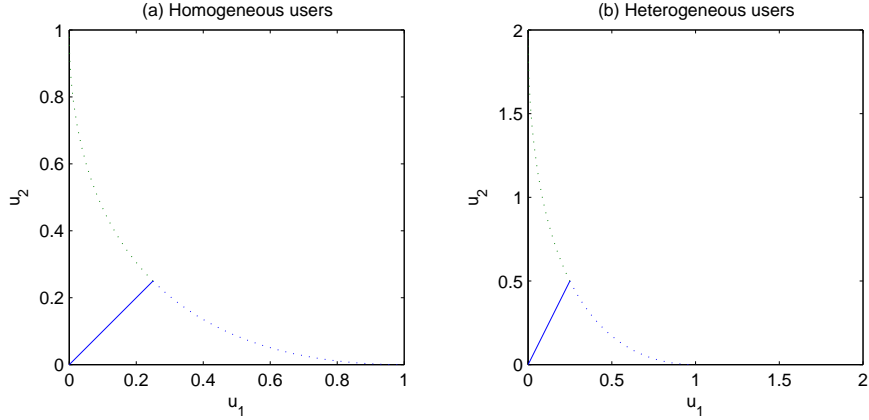}
\end{center}
\caption{Payoffs that can be achieved by the manager who observes
only the aggregate probability. (a) Homogeneous users with $k_1 =
k_2 =1$. (b) Heterogeneous users with $k_1 = 1$ and $k_2 = 2$.}
\end{figure}

Figure 5 shows the payoff profiles achieved with symmetric strategy
profiles, which can be implemented by the manager who observes the
aggregate probability.

\subsection{Limited Observability of the Users and Conjectural Equilibrium}
We now relax Requirement U and assume that user $i$ can observe only
the aggregate probability $(1 - p_0) \prod_{j \neq i} (1-p_j)$. Even
though the users do not know the exact form of the intervention
function of the manager, they are aware of the dependence of $p_0$
on their transmission probabilities and try to model this dependence
based on their observations $(p_i, (1 - p_0) \prod_{j \neq i}
(1-p_j))$. Specifically, user $i$ builds a conjecture function $f_i:
[0,1] \rightarrow [0,1]$, which means that user $i$ conjectures that
the value of $(1 - p_0) \prod_{j \neq i} (1-p_j)$ will be $f_i(p_i)$
if he chooses $p_i$. The equilibrium concept appropriate in this
context is \emph{conjectural equilibrium} first introduced by Hahn
\cite{hahn}.

\newtheorem{def5}[def1]{Definition}
\begin{def5}
A strategy profile $\hat{\mathbf{p}}$ and a profile of conjectures
$(f_1, \ldots, f_n)$ constitutes a \emph{conjectural equilibrium} of
the contention game with intervention $g$ if
\begin{eqnarray}
k_i \hat{p}_i \, f_i(\hat{p}_i) \geq k_i p_i \, f_i(p_i) \quad
\textrm{for all } p_i \in P_i
\end{eqnarray}
and
\begin{eqnarray}
f_i(\hat{p}_i) = (1-g(\hat{\mathbf{p}})) \prod_{j \neq i}
(1-\hat{p}_j)
\end{eqnarray}
for all $i \in N$.
\end{def5}
The first condition states that $\hat{p}_i$ is optimal given user
$i$'s conjecture $f_i$, and the second condition says that its
conjecture is consistent with its observation. It can be seen from
this definition that the conjectural equilibrium is a generalization
of Nash equilibrium in that any Nash equilibrium is a conjectural
equilibrium with every user holding the correct conjecture given
others' strategies. On the other hand, it is quite general in some
cases, and in the game we consider, for any strategy profile
$\hat{\mathbf{p}} \in P$, there exists a conjecture profile $(f_1,
\ldots, f_n)$ that constitutes a conjectural equilibrium. For
example, we can set $f_i(p_i) = (1-g(\hat{\mathbf{p}})) \prod_{j
\neq i} (1-\hat{p}_j)$ if $p_i = \hat{p}_i$ and 0 otherwise.

Since the TRD-based intervention function $g^*$ is linear in each
$p_i$, it is natural for the users to adopt a conjecture function of
the linear form. Let us assume that conjecture functions are of the
following trimmed linear form:
\begin{eqnarray}
f_i(p_i) = \left[ a_i - b_i p_i \right]_0^1
\end{eqnarray}
for some $a_i, b_i > 0$.

We say that a conjecture function $f_i$ is \emph{linearly
consistent} at $\hat{\mathbf{p}}$ if it is locally correct up to the
first derivative at $\hat{\mathbf{p}}$, i.e., $f_i(\hat{p}_i) =
(1-g(\hat{\mathbf{p}})) \prod_{j \neq i} (1-\hat{p}_j)$ and
$f'_i(\hat{p}_i) = - \frac{\partial g(\hat{\mathbf{p}})}{\partial
p_i} \prod_{j \neq i} (1-\hat{p}_j)$. Since the TRD-based
intervention function $g^*$ is linear in each $p_i$, the conjecture
function $f_i^*(p_i) \triangleq g^*(p_i, \tilde{\mathbf{p}}_{-i})$
is linearly consistent at $\tilde{\mathbf{p}}$, and
$\tilde{\mathbf{p}}$ and $(f_1^*,\ldots, f_n^*)$ constitutes a
conjectural equilibrium. Therefore, as long as the users use
linearly consistent conjectures, limited observability of the users
does not affect the final outcome. To build linearly consistent
conjectures, however, the users need to experiment and collect data
using local deviations from the equilibrium point in a repeated play
of the Stackelberg contention game. A loss in performance may result
during this learning phase.

\section{Illustrative Results}
\subsection{Homogeneous Users}
We assume that the users are homogeneous with $k_i = 1$ for all $i
\in N$. Given a transmission probability profile ${\mathbf{p}}$, the
system utilization ratio can be defined as the probability of
successful transmission in a given slot
\begin{eqnarray}
\tau(\mathbf{p}) = \sum_{i \in N} p_i \prod_{j \neq i} ( 1 - p_j).
\end{eqnarray}
Note that the maximum system utilization ratio is 1, which occurs
when only one user transmits with probability 1 while others never
transmit. Table 1 shows the individual payoffs and the system
utilization ratios for the number of users 3, 10, and 100 when the
manager implements the target at the symmetric efficient strategy
profile $\tilde{\mathbf{p}} = (\frac{1}{n}, \ldots, \frac{1}{n})$.

\begin{center}
\begin{tabular}{|c|c|c|}
\hline
$n$ & Individual Payoff & System Utilization Ratio \\
\hline
3 & 0.14815 & 0.44444 \\
10 & 0.03874 & 0.38742 \\
100 & 0.00370 & 0.36973 \\
\hline
\end{tabular}

\medskip Table 1. Individual payoffs and system utilization ratios with
homogeneous users
\end{center}
We can see that packets are transmitted in approximately 37\% of the
slots with a large number of users even if there is no explicit
coordination among the users. The system utilization of our model
converges to $1/e \approx 36.8\%$ as $n$ goes to infinity, which
coincides with the maximal throughput of a slotted Aloha system with
Poisson arrivals and an infinite number of users \cite{bert}. But in
our model users maintain their selfish behavior, and we do not use
any feedback information on the channel state.

\subsection{Heterogeneous Users}
We now consider users with difference valuations. Specifically, we
assume that $k_i = i$ for $i = 1, \ldots, n$. We will consider three
targets: $\tilde{\mathbf{p}}^1 = (1, \ldots, n)/\sum_{i=1}^n i$,
$\tilde{\mathbf{p}}^2 = (\frac{1}{n}, \ldots, \frac{1}{n})$, and
$\tilde{\mathbf{p}}^3$ with which
$\tilde{u}_i(\tilde{\mathbf{p}}^3;g^*) =
\tilde{u}_j(\tilde{\mathbf{p}}^3;g^*)$ for all $i,j$.
$\tilde{\mathbf{p}}^1$ assigns a higher transmission probability to
a user with a higher valuation. $\tilde{\mathbf{p}}^2$ treats all
the users equally regardless of their valuations.
$\tilde{\mathbf{p}}^3$ is egalitarian in that it yields the same
individual payoff to every user, which implies that a user with a
low valuation is assigned a higher transmission probability.

\begin{center}
\begin{tabular}{|c|c|c|c|c|c|c|c|}
\hline
& & Average & Aggregate
& Standard & System & Nash & Generalized\\
Target & $n$ & Individual & Payoff
& Deviation of & Utilization & Product & Nash\\
&  & Payoff &  &
Payoffs & Ratio & & Product\\
\hline
& 3 & 0.38889 & 1.16667 & 0.32710 & 0.47222 & 1.28601e-2 & 2.48073e-3 \\
$\tilde{\mathbf{p}}^1$ & 10 & 0.28048 & 2.80481 & 0.24643 & 0.39384 & 3.40193e-9 & 4.57497e-30\\
& 100 & 0.24855 & 24.85466 & 0.22189 & 0.37034 & 2.12632e-98 & $\approx$ 0\\
\hline
& 3 & 0.29630 & 0.88889 & 0.12096 & 0.44444 & 1.95092e-2 & 1.14183e-3 \\
$\tilde{\mathbf{p}}^2$ & 10 & 0.21308 & 2.13081 & 0.11127 & 0.38742 & 2.76432e-8 & 4.83117e-34\\
& 100 & 0.18671 & 18.67135 & 0.10673 & 0.36973 & 5.73364e-86 & $\approx$ 0\\
\hline
& 3 & 0.25133 & 0.75400 & 0 & 0.46078 & 1.58765e-2 & 2.52064e-4\\
$\tilde{\mathbf{p}}^3$ & 10 & 0.13753 & 1.37533 & 0 & 0.40283 & 2.42148e-9 & 4.09682e-48\\
& 100 & 0.07303 & 7.30337 & 0 & 0.37885 & 2.25070e-114 & $\approx$ 0\\
\hline
\end{tabular}

\medskip Table 2. Average individual payoffs, aggregate payoffs,
standard deviations of individual payoffs, system utilization
ratios, Nash products, and generalized Nash products with
heterogeneous users
\end{center}

Table 2 shows that a tradeoff between efficiency (measured by the
sum of payoffs) and equity exists when users are heterogeneous. A
higher aggregate payoff is achieved when users with high valuations
are given priority. At the same time, it limits access by users with
low valuations, which increases variations in individual payoffs.
Also, the results in Table 2 are consistent with that
$\tilde{\mathbf{p}}^2$ is a Nash bargaining solution and that
$\tilde{\mathbf{p}}^1$ is a nonsymmetric Nash bargaining solution
with weights equal to valuations.

\section{Conclusion}
We have analyzed the problem of multiple users who share a common
communication channel. Using the game theory framework, we have
shown that selfish behavior is likely to lead to a network collapse.
However, full system utilization requires coordination among users
using explicit message exchanges, which may be impractical given the
distributed nature of wireless networks. To achieve a better
performance without coordination schemes, users need to sustain
cooperation. We provide incentives for selfish users to limit their
access to the channel by introducing an intervention function of the
network manager. With TRD-based intervention functions, the manager
can implement any outcome of the contention game as a Stackelberg
equilibrium. We have discussed the amount of information required
for implementation, and how the various kinds of relaxations of the
requirements affect the outcome of the Stackelberg contention game.

Our approach of using an intervention function to improve network
performance can be applied to other situations in wireless
communications. Potential applications of the idea include
sustaining cooperation in multi-hop networks and limiting the attack
of adversary users. An intervention function may be designed to
serve as a coordination device in addition to providing selfish
users with incentives to cooperate. Finally, designing a protocol
that enables users to play the role of the manager in a distributed
manner will be critical to ensure that our approach can be adopted
in completely decentralized communication scenarios, where no
manager is present.

\appendix
\section{Proof of Proposition 3}
Recall $h(\mathbf{p}) = \frac{p_1}{\tilde{p}_1} + \cdots +
\frac{p_n}{\tilde{p}_n} - n$ used to define $g^*(\mathbf{p})$. We
examine whether a strategy profile $\hat{\mathbf{p}}$ with
$\hat{p}_i < 1$ for all $i \in N$ constitutes a Nash equilibrium of
$\Gamma_{g^*}$ by considering four cases on the value of
$h(\hat{\mathbf{p}})$.

\noindent \textbf{Case 1.} $h(\hat{\mathbf{p}}) < 0$.

Let $\epsilon = - h(\hat{\mathbf{p}}) > 0$. If user $i$ changes its
transmission probability from $\hat{p}_i$ to $\hat{p}_i + \epsilon$,
then its payoff increases because $p_0$ is still zero. Hence
$\hat{\mathbf{p}}$ cannot be a Nash equilibrium if
$h(\hat{\mathbf{p}}) < 0$.

\noindent \textbf{Case 2.} $h(\hat{\mathbf{p}}) = 0$.

Consider arbitrary user $i$. If it deviates to $p_i < \hat{p}_i$,
$p_0$ is still zero and $\tilde{u}_i$ decreases. $\tilde{u}_i(p_i,
\hat{\mathbf{p}}_{-i})$ is differentiable and strictly concave on
$p_i > \hat{p}_i$. Since $\frac{d \tilde{u}_i}{d p_i} = (k_i
\prod_{j \neq i} (1-\hat{p}_j)) (1 + n - \sum_{j \neq i}
\frac{\hat{p}_j}{\tilde{p}_j} - 2 \frac{p_i}{\tilde{p}_i})$, $k_i
> 0$ and $\hat{p}_i < 1$ for all $i$,
\begin{eqnarray}
sign \left(\frac{d \tilde{u}_i}{d p_i} \Bigg|_{p_i = \hat{p}_i}
\right) &=& sign \left(1 + n - \sum_{j \neq i}
\frac{\hat{p}_j}{\tilde{p}_j} - 2 \frac{\hat{p}_i}{\tilde{p}_i} \right)\\
&=& sign \left(1 + n - \sum_{j = 1}^n
\frac{\hat{p}_j}{\tilde{p}_j} - \frac{\hat{p}_i}{\tilde{p}_i} \right)\\
&=& sign \left(1 - \frac{\hat{p}_i}{\tilde{p}_i} \right).
\end{eqnarray}
There is no gain for user $i$ from deviating to any $p_i >
\hat{p}_i$ if and only if $\frac{d \tilde{u}_i}{d p_i}|_{p_i =
 \hat{p}_i} \leq 0$, which is equivalent to $\hat{p}_i \geq
\tilde{p}_i$. For $\hat{\mathbf{p}}$ to be a Nash equilibrium, we
need $\hat{p}_i \geq \tilde{p}_i$ for all $i = 1, \ldots, n$. To
satisfy $h(\hat{\mathbf{p}}) = 0$, all inequalities should be
equalities. Hence, only $\hat{\mathbf{p}} = \tilde{\mathbf{p}}$ is a
Nash equilibrium among $\hat{\mathbf{p}}$ such that
$h(\hat{\mathbf{p}}) = 0$.

\noindent \textbf{Case 3.} $0 < h(\hat{\mathbf{p}}) < 1$.

Since $\tilde{u}_i \geq 0$, there is no gain for user $i$ to deviate
to $p_i$ such that $h(p_i, \hat{\mathbf{p}}_{-i}) \geq 1$. If there
is a gain from deviation to $p_i$ such that $h(p_i,
\hat{\mathbf{p}}_{-i}) < 0$, then there is another profitable
deviation $p'_i$ such that $h(p'_i, \hat{\mathbf{p}}_{-i}) = 0$ by
using the argument of Case 1. Therefore, we can restrict our
attention to deviations $p_i$ that lead to $0 \leq h(p_i,
\hat{\mathbf{p}}_{-i}) < 1$. At such a deviation by user $i$,
\begin{eqnarray}
\tilde{u}_i(p_i,\hat{\mathbf{p}}_{-i}) = k_i \prod_{j \neq i}
(1-\hat{p}_j) p_i \left(1 + n - \sum_{j \neq i}
\frac{\hat{p}_j}{\tilde{p}_j} - \frac{p_i}{\tilde{p}_i} \right).
\end{eqnarray}
$\hat{p}_i$ is best response to $\hat{\mathbf{p}}_{-i}$ if and only
if $\frac{d \tilde{u}_i}{d p_i} |_{p_i = \hat{p}_i} = 0$. Using the
first derivative given in Case 2, we obtain
\begin{eqnarray}
\frac{\hat{p}_i}{\tilde{p}_i} = 1 + n - \sum_{j = 1}^n
\frac{\hat{p}_j}{\tilde{p}_j} = 1 - h(\hat{\mathbf{p}}) < 1.
\end{eqnarray}
For $\hat{\mathbf{p}}$ to be a Nash equilibrium, the above
inequality should be satisfied for every $i$, which in turn implies
\begin{eqnarray}
\sum_{i = 1}^n \frac{\hat{p}_i}{\tilde{p}_i} < n,
\end{eqnarray}
and this contradicts to the initial assumption $h(\hat{\mathbf{p}})
> 0$. Therefore, there is no $\hat{\mathbf{p}}$ with $0 < h(\hat{\mathbf{p}}) < 1$
that constitutes a Nash equilibrium.

\noindent \textbf{Case 4.} $h(\hat{\mathbf{p}}) \geq 1$.

Since $\tilde{u}_i(\hat{\mathbf{p}}) = 0$ for every $i$, there is a
profitable deviation of user $i$ only if there exists $p_i \in (0,
\hat{p}_i)$ such that $h(p_i,\hat{\mathbf{p}}_{-i}) < 1$.
Equivalently, if setting $p_i = 0$ yields
$h(p_i,\hat{\mathbf{p}}_{-i}) \geq 1$, then there is no profitable
deviation of user $i$ from $\hat{p}_i$. Since
\begin{eqnarray}
h(0, \hat{\mathbf{p}}_{-i}) = \sum_{j \neq i}
\frac{\hat{p}_j}{\tilde{p}_j} - n,
\end{eqnarray}
$\hat{p}$ with $h(\hat{\mathbf{p}}) \geq 1$ is a Nash equilibrium if
and only if
\begin{eqnarray}
\sum_{j \neq i} \frac{\hat{p}_j}{\tilde{p}_j} - n \geq 1 \textrm{
for all $i = 1, \ldots, n$}.
\end{eqnarray}

\section{Proof of Proposition 4}
Consider $t = 1$. User $i$ chooses $\hat{p}_i^1$ to maximize
\begin{eqnarray}
\tilde{u}_i^1(p_i, \hat{\mathbf{p}}_{-i}^0) &=& k_i p_i \left(1 -
g^1(p_i,
\hat{\mathbf{p}}_{-i}^0) \right) \prod_{j \neq i} (1-\hat{p}_j^0)\\
&=& \left\{ \begin{array}{ll} k_i \prod_{j \neq i} (1-\hat{p}_j^0)
p_i & \textrm{if $p_i < \tilde{p}_i \left( 1 -
\frac{\hat{p}_n^0}{\tilde{p}_n}
+ \frac{\hat{p}_i^0}{\tilde{p}_i} \right)$},\\
k_i \prod_{j \neq i} (1-\hat{p}_j^0) p_i \left(2 -
\frac{\hat{p}_n^0}{\tilde{p}_n} + \frac{\hat{p}_i^0}{\tilde{p}_i} -
\frac{p_i}{\tilde{p}_i} \right) & \textrm{if $\tilde{p}_i \left( 1 -
\frac{\hat{p}_n^0}{\tilde{p}_n} +
\frac{\hat{p}_i^0}{\tilde{p}_i}\right) \leq p_i \leq \tilde{p}_i
\left( 2 - \frac{\hat{p}_n^0}{\tilde{p}_n} +
\frac{\hat{p}_i^0}{\tilde{p}_i}\right)$},\\
0 & \textrm{if $p_i > \tilde{p}_i \left( 2 -
\frac{\hat{p}_n^0}{\tilde{p}_n} + \frac{\hat{p}_i^0}{\tilde{p}_i}
\right)$}.
\end{array} \right.
\end{eqnarray}

If $0 \leq \frac{\hat{p}_n^0}{\tilde{p}_n} -
\frac{\hat{p}_i^0}{\tilde{p}_i} < 2$, the maximum is attained at
$\hat{p}_i^1$ that satisfies
\begin{eqnarray} \label{eq:aaa}
\frac{\hat{p}_i^1}{\tilde{p}_i} = 1 - \frac{1}{2} \left(
\frac{\hat{p}_n^0}{\tilde{p}_n} - \frac{\hat{p}_i^0}{\tilde{p}_i}
\right).
\end{eqnarray}
Notice that $\hat{p}_n^1 = \tilde{p}_n$.

If $\frac{\hat{p}_n^0}{\tilde{p}_n} -
\frac{\hat{p}_i^0}{\tilde{p}_i} \geq 2$, then $\tilde{u}_i^1(p_i,
\hat{\mathbf{p}}_{-i}^0) = 0$ for all $p_i \geq 0$. Since any $p_i$
is a best response in this case, we assume that $\hat{p}_i^1 =
\hat{p}_i^0$.\footnote{If we assume that $\hat{p}_i^1$ is chosen
according to (\ref{eq:aaa}), we do not need the assumption that for
each $i$ either $\frac{\hat{p}_n^0}{\tilde{p}_n} -
\frac{\hat{p}_i^0}{\tilde{p}_i} < 2$ or
$\frac{\hat{p}_i^0}{\tilde{p}_i} \leq 1$ in the proposition.}

Consider $t = 2$. First, consider user $i$ such that
$\frac{\hat{p}_n^0}{\tilde{p}_n} - \frac{\hat{p}_i^0}{\tilde{p}_i} <
2$. Since $\frac{\hat{p}_n^1}{\tilde{p}_n} -
\frac{\hat{p}_i^1}{\tilde{p}_i} = \frac{1}{2}
\left(\frac{\hat{p}_n^0}{\tilde{p}_n} -
\frac{\hat{p}_i^0}{\tilde{p}_i} \right)$, $0 \leq
\frac{\hat{p}_n^1}{\tilde{p}_n} - \frac{\hat{p}_i^1}{\tilde{p}_i} <
2$. Using an analogous argument, we get
\begin{eqnarray}
\frac{\hat{p}_i^2}{\tilde{p}_i} = 1 - \frac{1}{2} \left(
\frac{\hat{p}_n^1}{\tilde{p}_n} - \frac{\hat{p}_i^1}{\tilde{p}_i}
\right) = 1 - \frac{1}{2^2} \left( \frac{\hat{p}_n^0}{\tilde{p}_n} -
\frac{\hat{p}_i^0}{\tilde{p}_i} \right).
\end{eqnarray}

Next consider user $i$ such that $\frac{\hat{p}_i^0}{\tilde{p}_i}
\leq 1$. Since $\frac{\hat{p}_n^1}{\tilde{p}_n} = 1$, we again have
$0 \leq \frac{\hat{p}_n^1}{\tilde{p}_n} -
\frac{\hat{p}_i^1}{\tilde{p}_i} < 2$ and the best response is given
by
\begin{eqnarray}
\frac{\hat{p}_i^2}{\tilde{p}_i} = 1 - \frac{1}{2} \left(
\frac{\hat{p}_n^1}{\tilde{p}_n} - \frac{\hat{p}_i^1}{\tilde{p}_i}
\right) = 1 - \frac{1}{2} \left( \frac{\hat{p}_n^1}{\tilde{p}_n} -
\frac{\hat{p}_i^0}{\tilde{p}_i} \right).
\end{eqnarray}

Considering a general $t \geq 2$, we get
\begin{eqnarray}
\frac{\hat{p}_i^t}{\tilde{p}_i} = 1 - \frac{1}{2^t} \left(
\frac{\hat{p}_n^0}{\tilde{p}_n} - \frac{\hat{p}_i^0}{\tilde{p}_i}
\right)
\end{eqnarray}
for user $i$ such that $\frac{\hat{p}_n^0}{\tilde{p}_n} -
\frac{\hat{p}_i^0}{\tilde{p}_i} < 2$ and
\begin{eqnarray}
\frac{\hat{p}_i^t}{\tilde{p}_i} = 1 - \frac{1}{2^{t-1}} \left(
\frac{\hat{p}_n^1}{\tilde{p}_n} - \frac{\hat{p}_i^0}{\tilde{p}_i}
\right)
\end{eqnarray}
for user $i$ such that $\frac{\hat{p}_i^0}{\tilde{p}_i} \leq 1$.
Taking limits as $t \rightarrow \infty$, we obtain the conclusions
of the proposition.

\section{Proof of Proposition 6}
Suppose that the users in the coalition $S = \{ i,j \}$ choose
$(p_i, p_j)$ instead of $(\tilde{p}_i, \tilde{p}_j)$. Then
\begin{eqnarray}
h(p_i, p_j, \tilde{\mathbf{p}}_{-S}) = \frac{p_i}{\tilde{p}_i} +
\frac{p_j}{\tilde{p}_j} + ( n - 2 ) - n = \frac{p_i}{\tilde{p}_i} +
\frac{p_j}{\tilde{p}_j} - 2,
\end{eqnarray}
and
\begin{eqnarray}
\tilde{u}_i(p_i, p_j, \tilde{\mathbf{p}}_{-S}) = k_i \prod_{k \notin
S} (1-\tilde{p}_k) p_i ( 1 - p_j ) ( 1 - g^*(p_i, p_j,
\tilde{\mathbf{p}}_{-S})),\\
\tilde{u}_j(p_i, p_j, \tilde{\mathbf{p}}_{-S}) = k_j \prod_{k \notin
S} (1-\tilde{p}_k) p_j ( 1 - p_i ) ( 1 - g^*(p_i, p_j,
\tilde{\mathbf{p}}_{-S}) ).
\end{eqnarray}
Hence, $\tilde{\mathbf{p}}$ is coalition-proof with respect to $S$
if and only if there does not exist $(p_i, p_j) \in [0,1]^2$ such
that
\begin{eqnarray}
p_i ( 1 - p_j ) ( 1 - g^*(p_i, p_j, \tilde{\mathbf{p}}_{-S}) )
&\geq& \tilde{p}_i ( 1 - \tilde{p}_j ) \label{eq:useri},\\
p_j ( 1 - p_i ) ( 1 - g^*(p_i, p_j, \tilde{\mathbf{p}}_{-S}) )
&\geq& \tilde{p}_j ( 1 - \tilde{p}_i ) \label{eq:userj}
\end{eqnarray}
with at least one inequality strict.

First, notice that setting $p_i = \tilde{p}_i$ and $p_j \neq
\tilde{p}_j$ will violate one of the two inequalities. The
inequality for user $i$ will not hold if $p_j
> \tilde{p}_j$, and the one for user $j$ will not hold if $p_j
< \tilde{p}_j$. Hence, both $p_i \neq \tilde{p}_i$ and $p_j \neq
\tilde{p}_j$ are necessary to have both inequalities satisfied at
the same time. We consider four possible cases.

\noindent \textbf{Case 1.} $p_i < \tilde{p}_i$ and $p_j
> \tilde{p}_j$

Since $g^*(\cdot) \geq 0$, (\ref{eq:useri}) is violated.

\noindent \textbf{Case 2.} $p_i > \tilde{p}_i$ and $p_j <
\tilde{p}_j$

Equation (\ref{eq:userj}) is violated.

\noindent \textbf{Case 3.} $p_i < \tilde{p}_i$ and $p_j <
\tilde{p}_j$

Since $h(p_i, p_j, \tilde{\mathbf{p}}_{-S}) < 0$, $g^*(p_i, p_j,
\tilde{\mathbf{p}}_{-S}) = 0$. Hence, (\ref{eq:useri}) and
(\ref{eq:userj}) become
\begin{eqnarray}
p_i ( 1 - p_j ) &\geq& \tilde{p}_i ( 1 - \tilde{p}_j ),\\
p_j ( 1 - p_i ) &\geq& \tilde{p}_j ( 1 - \tilde{p}_i ).
\end{eqnarray}
We consider the contour curves of $p_i ( 1 - p_j )$ and $p_j ( 1 -
p_i )$ going through $(\tilde{p}_i, \tilde{p}_j)$ in the
$(p_i,p_j)$-plane. The slope of the contour curve of $p_i ( 1 - p_j
)$ at $(\tilde{p}_i, \tilde{p}_j)$ is $\frac{1 -
\tilde{p}_j}{\tilde{p}_i}$ and that of $p_j ( 1 - p_i )$ is
$\frac{\tilde{p}_j}{1 - \tilde{p}_i}$. There is no area of mutual
improvement if and only if
\begin{eqnarray}
\frac{1 - \tilde{p}_j}{\tilde{p}_i} \geq \frac{\tilde{p}_j}{1 -
\tilde{p}_i},
\end{eqnarray}
which is equivalent to $\tilde{p}_i + \tilde{p}_j \leq 1$.

\noindent \textbf{Case 4.} $p_i > \tilde{p}_i$ and $p_j
> \tilde{p}_j$

Since $h(p_i, p_j, \tilde{\mathbf{p}}_{-S}) > 0$, $g^*(p_i, p_j,
\tilde{\mathbf{p}}_{-S}) = h(p_i, p_j, \tilde{\mathbf{p}}_{-S})$ as
long as $\frac{p_i}{\tilde{p}_i} + \frac{p_j}{\tilde{p}_j} \leq 3$.
Hence, (\ref{eq:useri}) and (\ref{eq:userj}) become
\begin{eqnarray}
p_i ( 1 - p_j ) \left( 3 - \frac{p_i}{\tilde{p}_i} -
\frac{p_j}{\tilde{p}_j} \right) &\geq& \tilde{p}_i ( 1 - \tilde{p}_j ),\\
p_j ( 1 - p_i ) \left( 3 - \frac{p_i}{\tilde{p}_i} -
\frac{p_j}{\tilde{p}_j} \right) &\geq& \tilde{p}_j ( 1 - \tilde{p}_i
).
\end{eqnarray}
The slope of the contour curve of $p_i ( 1 - p_j ) \left( 3 -
\frac{p_i}{\tilde{p}_i} - \frac{p_j}{\tilde{p}_j} \right)$ at
$(\tilde{p}_i, \tilde{p}_j)$ is
\begin{eqnarray}
\frac{(1 - \tilde{p}_j) \left(3 - 2\frac{\tilde{p}_i}{\tilde{p}_i} -
\frac{\tilde{p}_j}{\tilde{p}_j} \right)}{\tilde{p}_i \left(3 +
\frac{1}{\tilde{p}_j} - \frac{\tilde{p}_i}{\tilde{p}_i} -
2\frac{\tilde{p}_j}{\tilde{p}_j}\right)} = 0,
\end{eqnarray}
and that of $p_j ( 1 - p_i ) \left( 3 - \frac{p_i}{\tilde{p}_i} -
\frac{p_j}{\tilde{p}_j}\right)$ is
\begin{eqnarray}
\frac{\tilde{p}_j \left(3 + \frac{1}{\tilde{p}_i} -
2\frac{\tilde{p}_i}{\tilde{p}_i} - \frac{\tilde{p}_j}{\tilde{p}_j}
\right)}{(1 - \tilde{p}_i) \left(3 - \frac{\tilde{p}_i}{\tilde{p}_i}
- 2\frac{\tilde{p}_j}{\tilde{p}_j}\right)} = + \infty.
\end{eqnarray}
Therefore, there is no $(p_i, p_j) > (\tilde{p}_i, \tilde{p}_j)$
that satisfies (\ref{eq:useri}) and (\ref{eq:userj}) at the same
time.

\section{Proof of Proposition 7}
The ``if'' part is trivial because a strategy profile that is
coalition-proof with respect to the grand coalition is Pareto
efficient. To establish the ``only if'' part, we will prove that if
for a given strategy profile there exists a coalition that can
improve the payoffs of its members then its deviation will not hurt
other users outside of the coalition, which shows that the original
strategy profile is not Pareto efficient.

Consider a strategy profile $\tilde{\mathbf{p}}$ and a coalition $S
\subset N$ that can improve upon $\tilde{\mathbf{p}}$ by deviating
from $\tilde{\mathbf{p}}_S$ to $\mathbf{p}_S$. Let $p_0 =
g^*(\mathbf{p}_S, \tilde{\mathbf{p}}_{-S})$ the transmission
probability of the manager after the deviation by coalition $S$.
Since choosing $\mathbf{p}_S$ instead of $\tilde{\mathbf{p}}_S$
yields higher payoffs to the members of $S$, we have
\begin{eqnarray} \label{eq:coal}
p_i (1-p_0) \prod_{j \in S \setminus \{i\}} (1 - p_j) \geq
\tilde{p}_i \prod_{j \in S \setminus \{i\}} (1 - \tilde{p}_j)
\end{eqnarray}
for all $i \in S$ with at least one inequality strict. We want to
show that the members not in the coalition $S$ do not get lower
payoffs as a result of the deviation by $S$, that is,
\begin{eqnarray}
(1-p_0) \prod_{j \in S} (1 - p_j) \geq \prod_{j \in S} (1 -
\tilde{p}_j).
\end{eqnarray}

Suppose $(1-p_0) \prod_{j \in S} (1 - p_j) < \prod_{j \in S} (1 -
\tilde{p}_j)$. We can see that $p_0 < 1$ and $0 < p_i < 1$ for all
$i \in S$ because the right-hand side of (\ref{eq:coal}) is strictly
positive. Combining this inequality with (\ref{eq:coal}) yields $p_i
> \tilde{p}_i$ for all $i \in S$, which implies $p_0 > 0$.

We can write $p_i = \tilde{p}_i + \epsilon_i$ for some $\epsilon_i >
0$ for $i \in S$. Then $p_0 = g^*(\mathbf{p}_S,
\tilde{\mathbf{p}}_{-S}) = \sum_{i \in S}
\frac{\epsilon_i}{\tilde{p}_i}$. (\ref{eq:coal}) can be rewritten as
\begin{eqnarray}
\tilde{p}_i \prod_{j \in S \setminus \{i\}} (1 - \tilde{p}_j) &\leq&
(\tilde{p}_i + \epsilon_i) (1-p_0) \prod_{j \in S \setminus \{i\}}
(1 - \tilde{p}_j - \epsilon_j)\\ &<& (\tilde{p}_i + \epsilon_i)
(1-p_0) \prod_{j \in S \setminus \{i\}} (1 - \tilde{p}_j)
\end{eqnarray}
for all $i \in S$. Simplifying this gives
\begin{eqnarray}
\frac{\epsilon_i}{\tilde{p}_i} > \frac{p_0}{1 - p_0}
\end{eqnarray}
for all $i \in S$. Summing these inequalities up over $i \in S$, we
get
\begin{eqnarray}
p_0 = \sum_{i \in S} \frac{\epsilon_i}{\tilde{p}_i} > |S|
\frac{p_0}{1 - p_0}
\end{eqnarray}
where $|S|$ is the number of the members in $S$. This inequality
simplifies to $p_0 < 1 - |S| \leq 0$, which is a contradiction.

\end{document}